\documentclass{aip-cp}

\usepackage[10pt]{type1ec}  
\usepackage[T1]{fontenc}
\usepackage{CJKutf8}
\usepackage{amssymb}
\usepackage[overlap, CJK]{ruby}
\usepackage{CJKulem}
\newenvironment{SChinese}{%
  \CJKfamily{gbsn}%
  \CJKtilde
  \CJKnospace}{}

\usepackage[numbers]{natbib}
\usepackage{rotating}
\usepackage{graphicx}
\usepackage{color}

\usepackage{mathrsfs}
\usepackage{amssymb}
\usepackage{bm}
\usepackage{longtable}
\usepackage{times}

\renewcommand\sout{\bgroup \color{red} \ULdepth=-.5ex \ULset}

\newcommand{\ba}{\begin{array}}
\newcommand{\ea}{\end{array}}

\newcommand{\ntrans}{n_{\rm trans}}

\newcommand{\etrans}{\varepsilon_{\rm trans}}
\newcommand{\Mmax}{M_{\rm max}}

\newcommand{\Msolar}{{\rm M}_{\odot}}

\newcommand{\cQMsq}{c^2_{\rm QM}}

\newcommand{\al}{\alpha}

\newcommand{\De}{\Delta}
\newcommand{\ep}{\varepsilon}
\newcommand{\eps}{\epsilon}

\newcommand{\La}{\Lambda}

\begin{document}

\title{Observability of sharp phase transitions in neutron stars}
\author[aff1,aff2]{Sophia Han (\begin{CJK}{UTF8}{}\begin{SChinese}韩 君\end{SChinese}\end{CJK})\noteref{note1}}

\affil[aff1]{Department of Physics and Astronomy, Ohio University, Athens, OH~45701, USA}
\affil[aff2]{Department of Physics, University of California Berkeley, Berkeley, CA~94720, USA}

\authornote[note1]{E-mail address: sjhan@berkeley.edu}

\maketitle

\begin{abstract}
With central densities as high as 5-10 times the nuclear saturation density, neutron stars exhibit extreme conditions that cannot be observed elsewhere. They are ideal astrophysical laboratories for probing the composition and properties of cold, ultra-dense matter. We shall discuss taking into account currently available data from observation, how to reveal possible sharp phase transitions in dense neutron star cores.

\end{abstract}

\section{INTRODUCTION}
\label{sec:intro}

We observe neutron stars by detecting photons, neutrinos and gravitational waves emitted from them. The groundbreaking discoveries of two-solar mass pulsars \cite{Demorest:2010bx,Antoniadis:2013pzd} renewed the lower bound on neutron star maximum mass, which impose stringent constraint on the theoretical models. The gravitational-wave (GW) emission from a binary neutron star (BNS) merger was first detected in August 2017 \cite{LIGO:2017qsa}, followed by its electromagnetic counterparts seen by many observatories all over the world. It was confirmed that information on the tidal deformability of neutron stars, a physical quantity that describes to what extent stars are deformed in response to an external tidal field and is sensitive to the radius, can be extracted from the pre-merger gravitational-wave signal.
Other remarkable observational advances include, continued monitoring of the thermal radiation from accreting and isolated neutron stars which provides the evidence for the existence of a neutron star crust, accurate measurements of changes in the spin rate of fast-rotating pulsars inferring the presence of some superfluid component in their interior, and the serendipitous observation of neutrinos and visible light from the Type II supernova SN 1987A in the Large Magellanic Cloud.

Global aspects of neutron stars are determined by the equations of state (EoS) of dense matter encountered in their interiors, which relates the energy density to the pressure; their thermal and spin evolution, as well as the gravitational-wave and neutrino spectra are instead highly sensitive to the material properties such as opacity, viscosity, elasticity, superfluidity, and also to the neutrino-matter interaction, neutrino oscillation, et cetera. These properties differ significantly among various forms of matter because of the low-energy degrees of freedom and their interactions. Taking into account that there are many choices, it will be necessary to combine all available observational data in a comprehensive framework to discriminate them. Strong first-order phase transitions in particular, if realized in the neutron star cores, can lead to some maximally distinguishable signatures.

\section{GLOBAL ASPECTS OF STABLE HYBRID STARS}
\label{sec:eos} 

The global structure of compact stars, including properties such as the mass, radius, tidal deformability, and the moment of inertia, are obtained for a realistic EoS by numerically solving Einstein equations of hydrostatic equilibrium, the Tolman-Oppenheimer-Volkov (TOV) equations \cite{Tolman:1939jz,Oppenheimer:1939ne}. Due to poorly constrained short-distance behavior of two- and three-nucleon interaction, there are large uncertainties in the nuclear matter EoS above 1-2 times saturation density, leading to variations in predicted radii and maximum masses of neutron stars. In the densest inner cores of neutron stars, exotic matter such as hyperons, meson condensates, or deconfined quarks may appear and even dominate. There are phenomenological models that describe possible novel phases, limited by the non-perturbative nature of Quantum Chromodynamics (QCD) at pertinent densities that prevents rigorous predictions.

So far we have two best, independent constraints from observation on the ``stiffness'' of equations of state, which characterizes how fast the pressure $p$ is rising with energy density $\ep$. 
Stiff EoSs in general give rise to larger neutron stars with higher maximum masses. EoSs that are too soft have been eliminated by the discovery of massive pulsars, whereas the pre-merger gravitational-wave signal indicates that too stiff EoSs correlated with large tidal deformation seem to be inconsistent with observation. 

\begin{figure*}[htb]
\parbox{0.48\hsize}{
\includegraphics[width=\hsize]{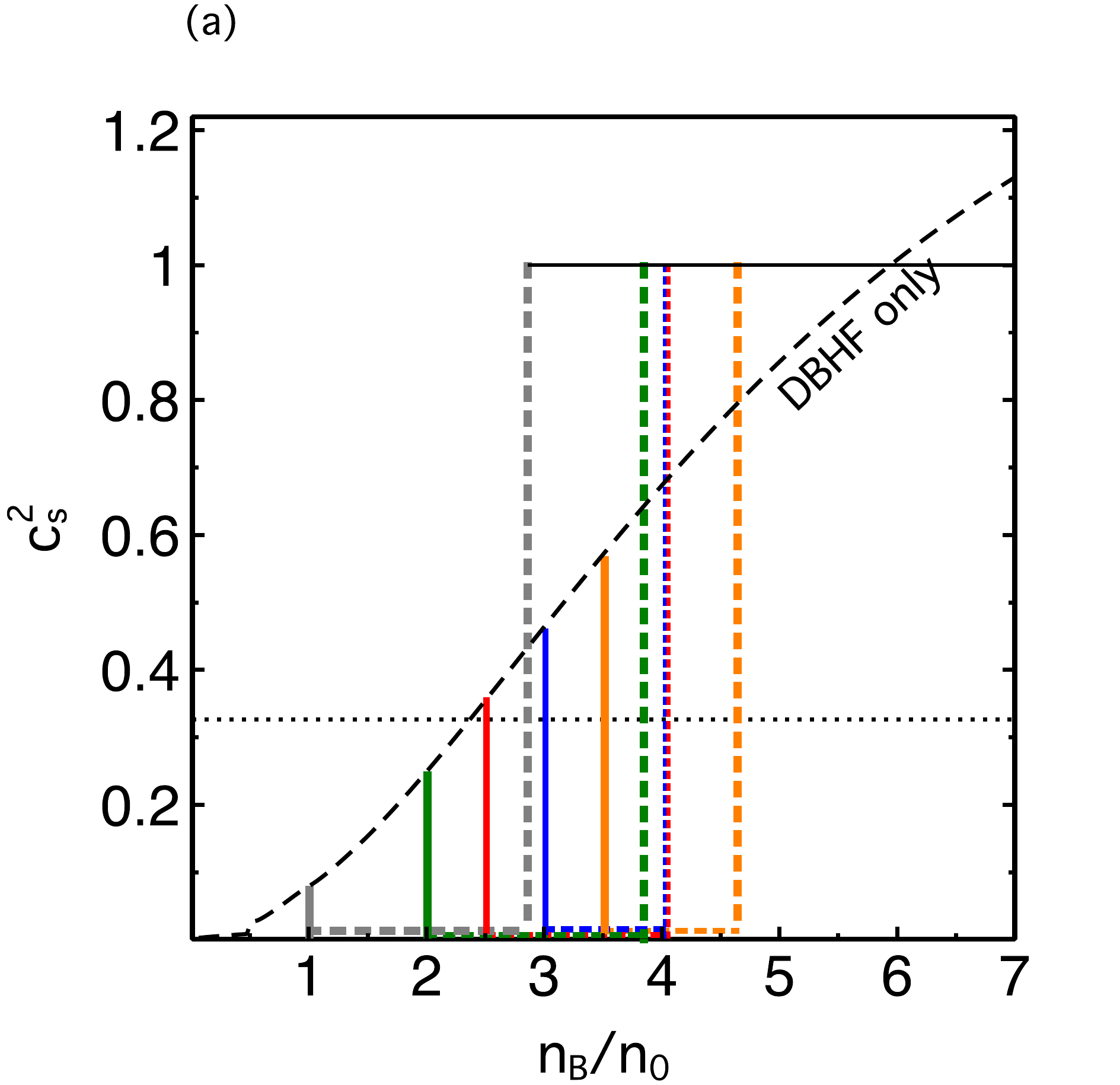}\\
}\parbox{0.5\hsize}{
\includegraphics[width=\hsize]{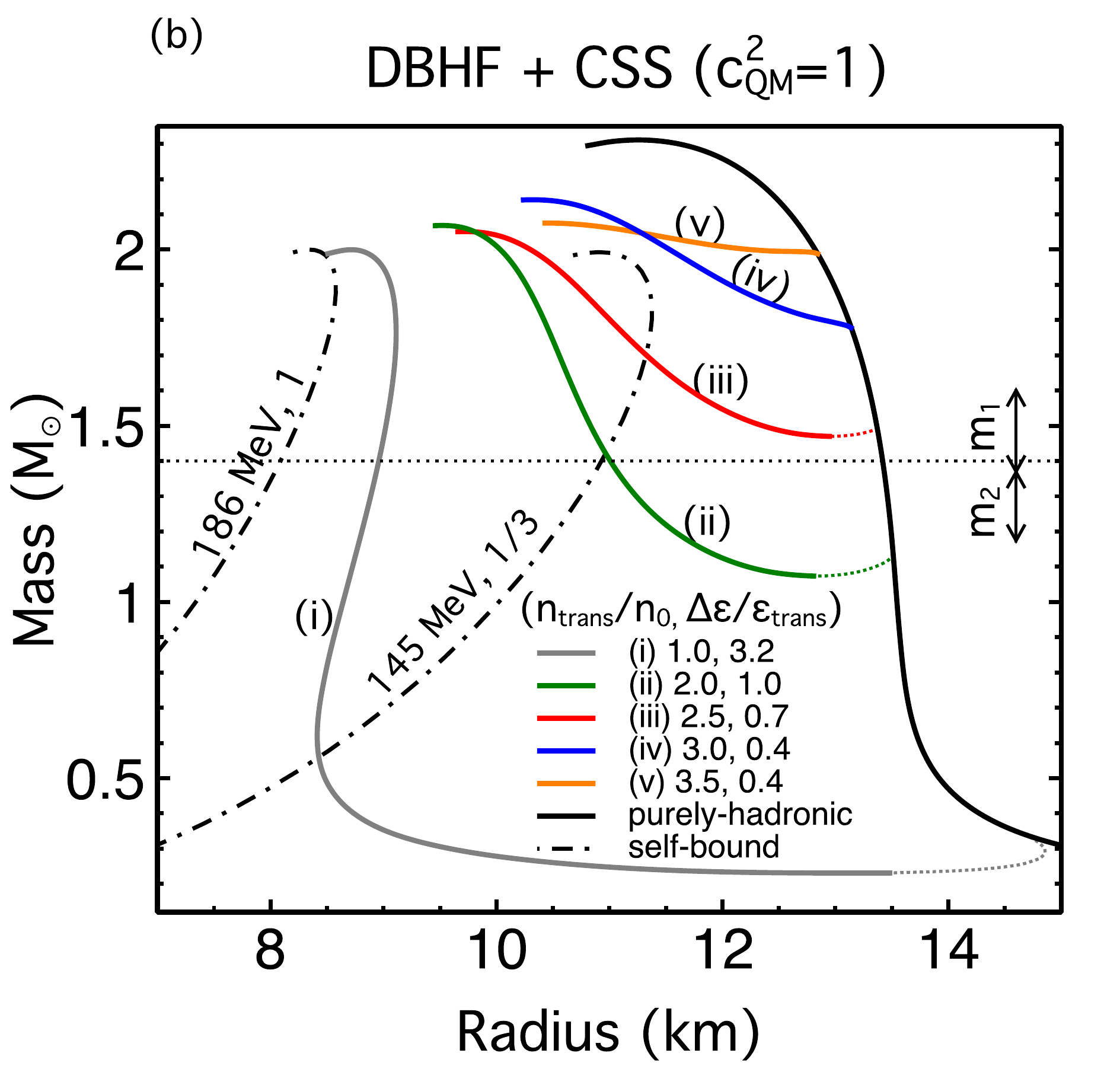}\\
}
\caption{Exemplary hybrid EoSs with a sharp phase transition from hadronic matter (DBHF) to quark matter (CSS). Panel (a): the sound-speed-squared as a function of the baryon number density for DBHF hadronic EoS (black dashed), with the density range for various sharp transitions enclosed by colored vertical lines. For each set of two vertical lines with the same color, the left one (solid) starting from hadronic curve down to zero marks the onset density of phase transition in hadronic matter $\ntrans$, and the one to the right (dotted) jumping to maximal $c_s^2=1$ refers to the density in quark matter just above phase transition. Note that $1/c_s^2$ encounters a singularity at $\ntrans$, not a finite discontinuity. Two horizontal lines represent the causality limit $c_s^2=1$ and the conformal limit $c_s^2=1/3$. Panel (b): corresponding mass-radius relations for the selected EoSs shown in panel (a). Self-bound strange quark stars (SQSs) are plotted for comparison, with two sets of parameters chosen such that $\Mmax=2\,\Msolar$ with $\cQMsq=1$ or 1/3. Two arrows labelled ``$m_1$'' and ``$m_2$'' stand for estimated ranges of the component mass in GW170817. } 
\label{fig:csq-mr-dbhf-css}
\end{figure*}

The tension between the existence of massive pulsars and the small tidal deformation inferred from the one observed merger (and also the possibly small radii from X-ray observations) can be reconciled by: 
\begin{enumerate}
\item a rapid switch from a soft EoS to a stiff EoS over a narrow pressure range in normal hadronic matter (the ``minimal scenario''), as indicated from e.g. computations of chiral effective field theory (chEFT)~\cite{Hebeler:2010jx} and quantum Monte Carlo (QMC) methods~\cite{Gandolfi:2011xu}, or\\
\item a crossover region of hadron-quark duality (the ``crossover scenario'') presumably within 1.5-5 times saturation density, generally yielding simple mass-radius topology (as in e.g. quarkyonic matter or interpolated EoS~\cite{Masuda:2012ed,Fukushima:2015bda,Baym:2017whm,McLerran:2018hbz}), or\\ 
\item strong first-order phase transition(s) at some supranuclear densities, creating an intrinsic softening of the EoS due to the finite energy density discontinuity with sufficiently stiff quark matter above the transition.
\end{enumerate}

It is noteworthy that the speed of sound, $c_s=\sqrt{\partial p/\partial \varepsilon}$, is a direct measure of the stiffness model-independently, and its behavior differs drastically in the three scenarios aforementioned (apart from obeying the causality limit $c_s^2 \leq 1$ universally). 
Although at asymptotically high densities all EoSs should approach the QCD limit $c_s^2 \approx 1/3$ \cite{Borsanyi:2012cr,Kurkela:2014vha,Bedaque:2014sqa}, uncertainties of the speed of sound behavior at intermediate densities relevant to neutron star cores are still huge. It has been shown that to be compatible with the observational constraint $\Mmax\geq2\,\Msolar$, within normal nuclear matter scenario $c_s$ is possibly reaching values closer to the speed of light  at a few times nuclear saturation density; strong first-order transitions induce diminishing $c_s^2$ at the quark-hadron interface (see e.g.  dotted segments in FIGURE \ref{fig:csq-mr-dbhf-css} (a)) with a certain energy density gap, for which the allowed parameter space is larger if the sound speed in quark matter is higher (preferably $\cQMsq \gtrsim 0.5$ when the hadronic EoS is soft and $\cQMsq \gtrsim 0.4$ when the hadronic EoS is stiff \cite{Alford:2013aca,Alford:2015gna}); for smoothly continuous hadron-quark crossover transition, $c_s^2$ undergoes change in concavity at least twice. 

Below we shall elaborate on the ``sharp phase-transition'' scenario and demonstrate its observability with regard to measuring global, static quantities; given the uncertainty in estimates of the surface tension between hadronic and quark matter, sharp interface (Maxwell construction) is assumed instead of mixed phase (Gibbs construction).

\begin{figure*}[htb]
\parbox{0.52\hsize}{
\includegraphics[width=\hsize]{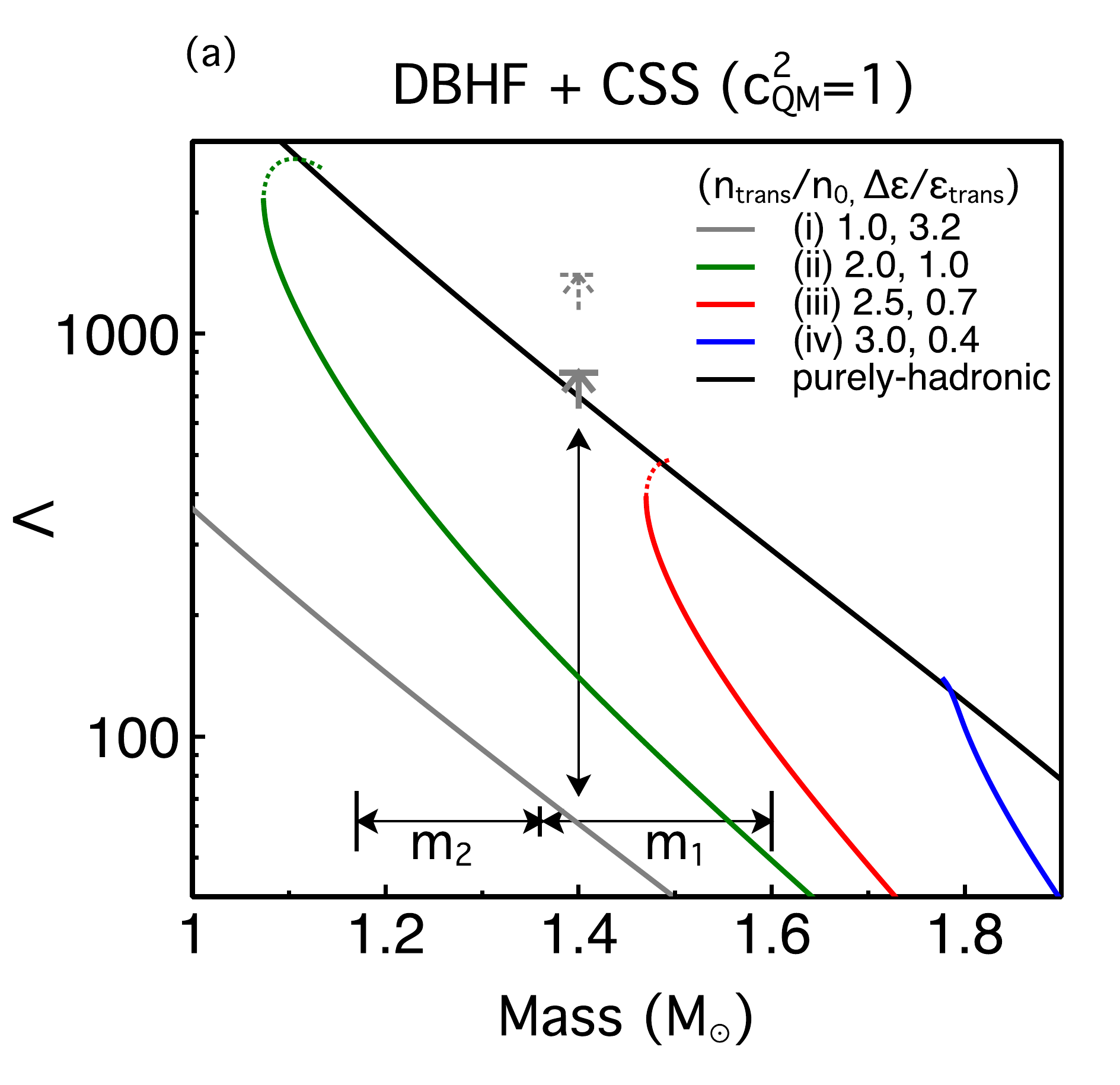}\\
}\parbox{0.5\hsize}{
\includegraphics[width=\hsize]{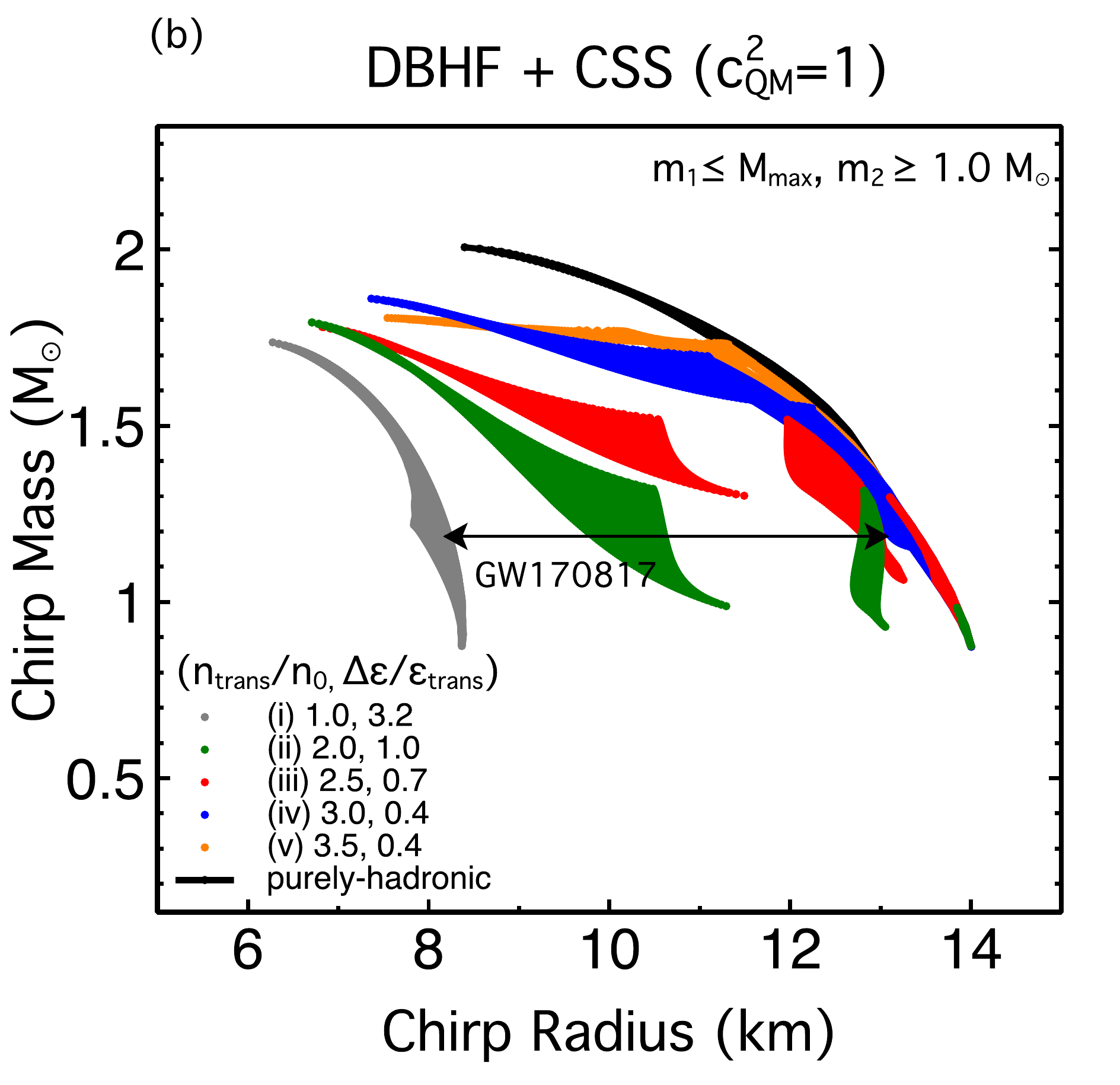}\\
}
\caption{Relevant binary neutron star (BNS) merger observables. Panel (a): the dimensionless tidal deformability as a function of mass $\La(M)$ for the same hybrid EoS parameters as in FIGURE~\ref{fig:csq-mr-dbhf-css}; solid (dotted) parts denote stable (unstable) configurations. Panel (b): the resulting chirp mass and ``chirp radius'' \cite{Wade:2014vqa} for all possible combinations of the component mass that satisfy $m_1\leq\Mmax$, $m_2\geq 1\,\Msolar$. The latest estimate for $\La$ in GW170817 is shown for comparison. The DBHF EoS is probably excluded by this single detection, however can be revived if a strong phase transition occurs in pre-merger NSs.
} 
\label{fig:m-Lam-dbhf-css}
\end{figure*}

We applied the generic ``constant-speed-of-sound'' (CSS) parameterization \cite{Alford:2013aca} of a first-order transition to a high-density phase, specifying the critical pressure at which the transition occurs, the strength of the transition, and the ``stiffness'' of the high-density phase. For hadronic matter, the Dirac-Brueckner-Hartree-Fock (DBHF) EoS~\cite{GrossBoelting:1998jg} is chosen for illustrative purposes. On the one hand, placing the constraint that NS maximum mass must be above $2\,\Msolar$ already rules out a considerable parameter space for quark matter \cite{Alford:2013aca,Alford:2015gna}; should in the future a pulsar heavier than $2\,\Msolar$ be detected\footnote{This is most likely the case very recently reported in \cite{Cromartie:2019kug}, where relativistic Shapiro delay measurements of PSR J0740+6620 revealed its mass $M=2.17_{-0.10}^{+0.11}\Msolar$ at $1-\sigma$ level.}, we anticipate even better limits on the phase transition properties. In addition, a reasonable upper bound on the radius of a massive pulsar inferred from multi-messenger observations holds the promise of ruling out not only nuclear matter EoSs that are too stiff, but also quark matter EoSs that are too soft. For instance, if it is confirmed that $R_{M=2.1\,\Msolar}\lesssim 12\,\rm{km}$ or $R_{M=2.0\,\Msolar}\lesssim 11.5 \,\rm{km}$, then soft quark matter with $\cQMsq\leq 1/3$ becomes incompatible immediately 
regardless of the feature of nuclear matter at lower densities \cite{Alford:2015gna}. On the other hand, allowing strong sharp phase transitions typically induce smaller tidal deformability of neutron stars compared to normal nuclear EoSs, resulting from sizable changes in both the radius and tidal Love number $k_2$ \cite{Han:2018mtj}.  

The most distinctive feature of hybrid EoSs with a sharp phase transition is the possible discontinuity in the mass-radius relation, as well as the drastic change in the tidal deformability from below to above the threshold for phase transition. If the phase transition is sufficiently strongly first-order (large enough $\De \ep$) and occurs at low enough density $\ntrans$, there can exist a hybrid star branch on the $M-R$ (and $\Lambda -M$) diagram that is disconnected from the normal hadronic branch (e.g., curves (i), (ii), (iii) in FIGURE~\ref{fig:csq-mr-dbhf-css}(b) and FIGURE~\ref{fig:m-Lam-dbhf-css}), the ``third-family'' of stars. If there is a second phase transition, then there might be ``fourth-family'' hybrid stars with even smaller radius and tidal deformability~\cite{Alford:2017qgh,Han:2018mtj}. Consequently, allowing phase transitions may rescue normal stiffer EoSs that were ruled out by the LIGO detection of GW170817 \cite{LIGO:2018wiz}.

The emergence of discontinuous mass-radius relation as well as multivalued tidal deformabilities is essential for distinguishing quark matter using global quantity measurements. These distinctive features could possibly be realized only in the strong sharp phase-transition scenario, unlike in the crossover or weak phase-transition case where exotic matter basically resembles normal nuclear matter in terms of static properties that are solely determined by the EoS, and becomes barely discernible. Note that self-bound strange quark stars (SQSs) possess unique mass-radius relation, and they form a separate family of stellar configurations.

Another promising global observable that can potentially help discriminate exotic matter is the moment of inertia, which enters into several well-known universal relations that are fairly insensitive to the EoSs if phase transitions are not taken into account. For example, the strong correlation between the dimensionless moment of inertia $\bar{I}= I/M^3$ and dimensionless tidal deformability $\La$ is found to be accurate within $1\%$~\cite{Yagi:2013awa,Yagi:2016bkt}, but with sequential first-order phase transitions into quark matter the deviation can be as large as $\sim9\%$ \cite{Han:2018mtj}. 

In summary, to ``detect'', or to infer exotic matter from measuring global, static quantities of neutron stars are probably very difficult or even unfeasible, unless there are strong first-order phase transitions that dramatically affect the radius and tidal deformability, and also challenge universal relations involving the moment of inertia.

\section{HINTS THAT NEUTRONS AND PROTONS ARE NOT ENOUGH}
\label{sec:obs} 

Neutron stars lose energy by neutrino emissions from the interior and photon emissions from the surface. In normal scenario of neutrons and protons, only massive enough neutron stars can possibly trigger fast cooling through direct Urca and are much colder, of which the threshold is controlled by the density dependence of symmetry energy in the EoS~\cite{Steiner:2004fi}. Besides, the existence of superfluidity quenches other cooling processes, but at the same time introduces additional cooling through the breaking and re-formation of Cooper pairs in close proximity to the superfluid critical temperature. Superfluidity also alters the specific heat that can affect the cooling efficiency (for a review, see e.g. \cite{Page:2013hxa}).

While excluding fast neutrino-emission mechanisms such as nucleonic direct Urca can still be compatible with current data for isolated cooling neutron stars, as long as Cooper-pair induced neutrino emissivity is suitably operating to accommodate certain sources, it is nearly impossible to avoid fast neutrino emission from observed cooling sources in transiently accreting systems~\cite{Brown:2017gxd,Heinke10,Wijnands:2017jsc}. Whether the enhanced cooling required is associated with exotic matter remains nevertheless unknown, which necessitate a further complete study quantifying all possible uncertainties. It is also intriguing that the estimated low to intermediate mass for the accreting neutron star in SAX J1808.4 3658~\cite{Morsink:2009wv} could hardly fit in standard models given its extremely low luminosity~\cite{Han:2017jaj}, which implies some exotic composition might be present in its core.

More evidence for the need of additional physics comes from the confrontation of r-mode instability window with accumulated X-ray data of spins and temperatures of neutron stars in low-mass X-ray binaries (LMXBs). Nascent neutron stars could spin as fast as allowed by the mass-shedding limit ($\sim$ 1\,kHz), at which the velocity of the surface equals that of an orbiting particle just above the surface. Observed young pulsars however, all spin with surprisingly low frequency around 10-100 Hz. One plausible explanation for this phenomenon is that r-modes of global, non-radial oscillations were driven into instability, which grow exponentially emitting gravitational waves that quickly spin down the star on a timescale of days to months after the neutron star was born~\cite{Andersson:2000mf}. The r-mode instability window is determined by the balance between the gravitational-wave radiation which is the driving force, and the viscous dissipation from particle scattering and transformation in dense matter which is the damping force. The ``minimal model'' based on hadronic matter only is insufficient to explain fast-rotating neutron stars we observe in LMXBs, irrespective of the specific EoS models~\cite{Haskell:2015iia,Kokkotas:2015gea}.

\section{PHASE-CONVERSION DISSIPATION (PCD) AT SHARP INTERFACES}
\label{sec:pcd}

Possible nonlinear saturation mechanisms that can saturate r-modes at tiny amplitudes compatible with observations $\al\lesssim 10^{-8}$ \cite{Mahmoodifar:2013quw,Mahmoodifar:2017ccc} would require either superfluidity~\cite{Haskell:2013hja} or phase conversion between hadrons and quarks~\cite{Alford:2014jha}. However, the former mechanism is restricted by superfluid critical temperatures, prohibiting it from working in warm stars; the latter comes with a range of simplifications (e.g. steady-state motion of the phase boundary) and is only effective for reasonably big quark cores, with superfluid effects being neglected at least close to the hadron/quark interface. Here we briefly review the basic concept of phase-conversion dissipation (PCD) for sharp transitions in neutron stars, and report a first estimate on its EoS dependence.

\begin{figure*}[htb]
\parbox{0.53\hsize}{
\includegraphics[width=\hsize]{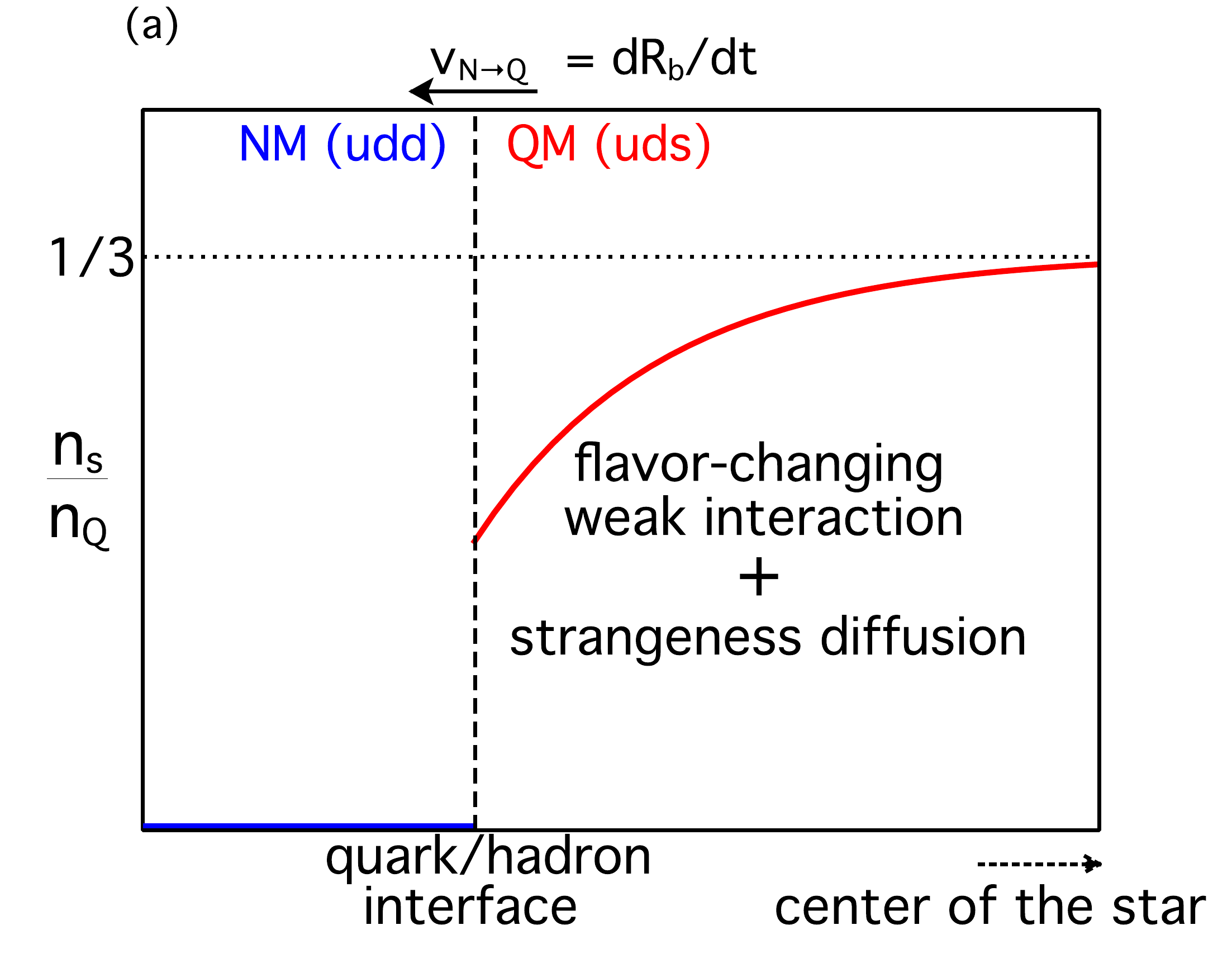}\\
}\parbox{0.5\hsize}{
\includegraphics[width=\hsize]{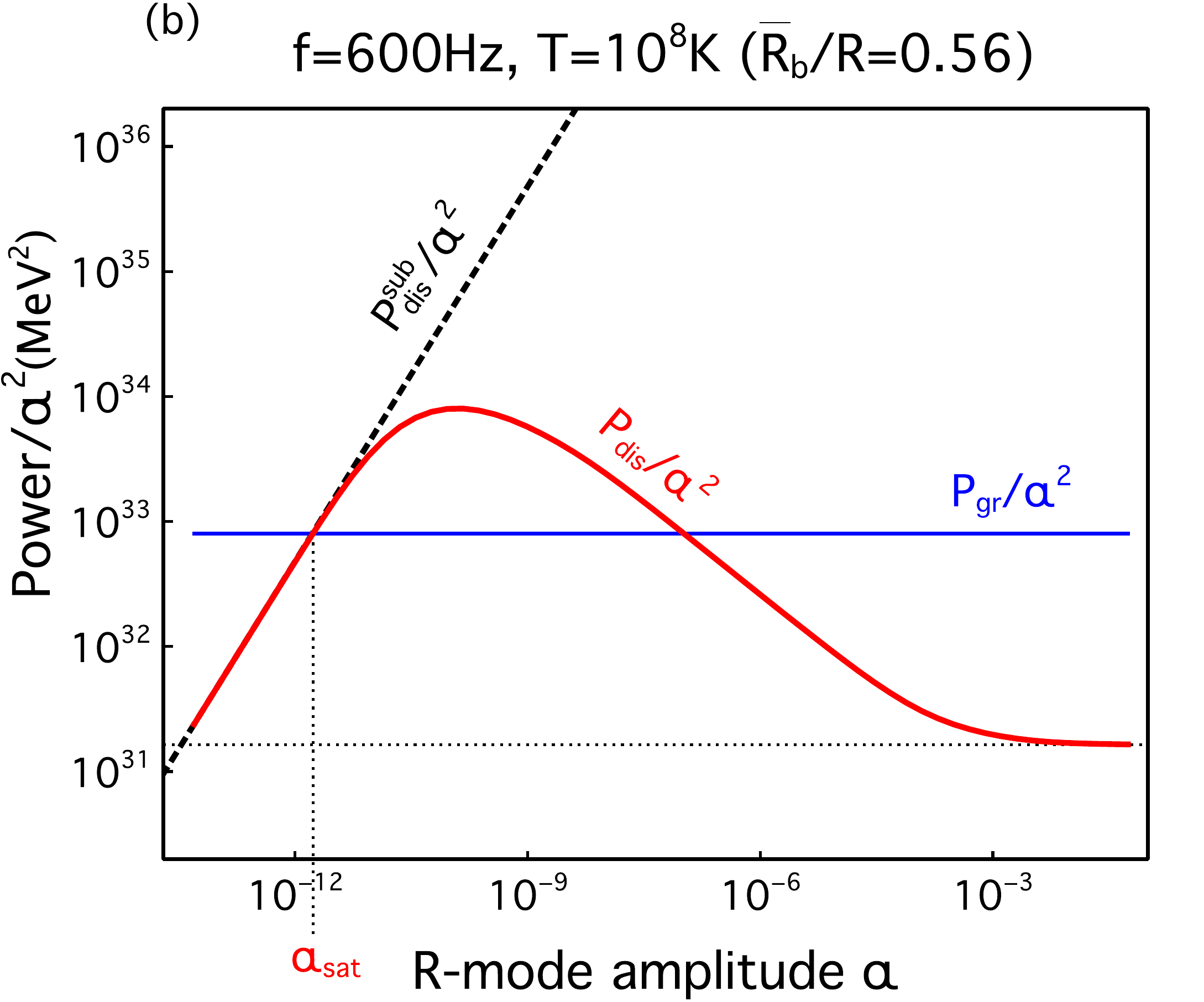}\\
}
\caption{Panel (a): schematic plot of the strangeness fraction profile close to the  interface where (for the half cycle shown) pure neutron matter is being converted into pure quark matter, as the boundary moving from the high-density side to the low-density side with velocity $v_{\rm N\to Q}=\rm {d R_{b}/d t}$ using steady-state approximation. Panel (b): dissipated power due to phase conversion (solid red curve) as a function of the r-mode amplitude $\al$ for a specific example hybrid star; adapted from \cite{Alford:2014jha}.
} 
\label{fig:ph-con-alpha}
\end{figure*}

Due to global density/pressure oscillations caused by e.g. r-modes, the flavor-changing process $d\leftrightarrow s$ is driven out of equilibrium. Finite rates of the weak interaction and flavor diffusion together lead to a phase lag in the system's response periodically which dissipates energy (see FIGURE \ref{fig:ph-con-alpha}(a) for a schematic), in analogous to the standard bulk viscosity which also exhibits a resonant behavior. Effectively, in the subthermal regime ($\De\mu\ll T \ll\mu$) the hadron/quark interface moves back and forth, converting hadrons into quarks and vice versa. The boundary velocity is determined by the diffusion coefficient for strangeness and the weak interaction rate, and because in nuclear matter diffusion is less efficient and weak interactions take more time to proceed, the quark-to-hadron half cycle dominates the dissipated energy.

\begin{figure*}[htb]
\parbox{0.53\hsize}{
\includegraphics[width=\hsize]{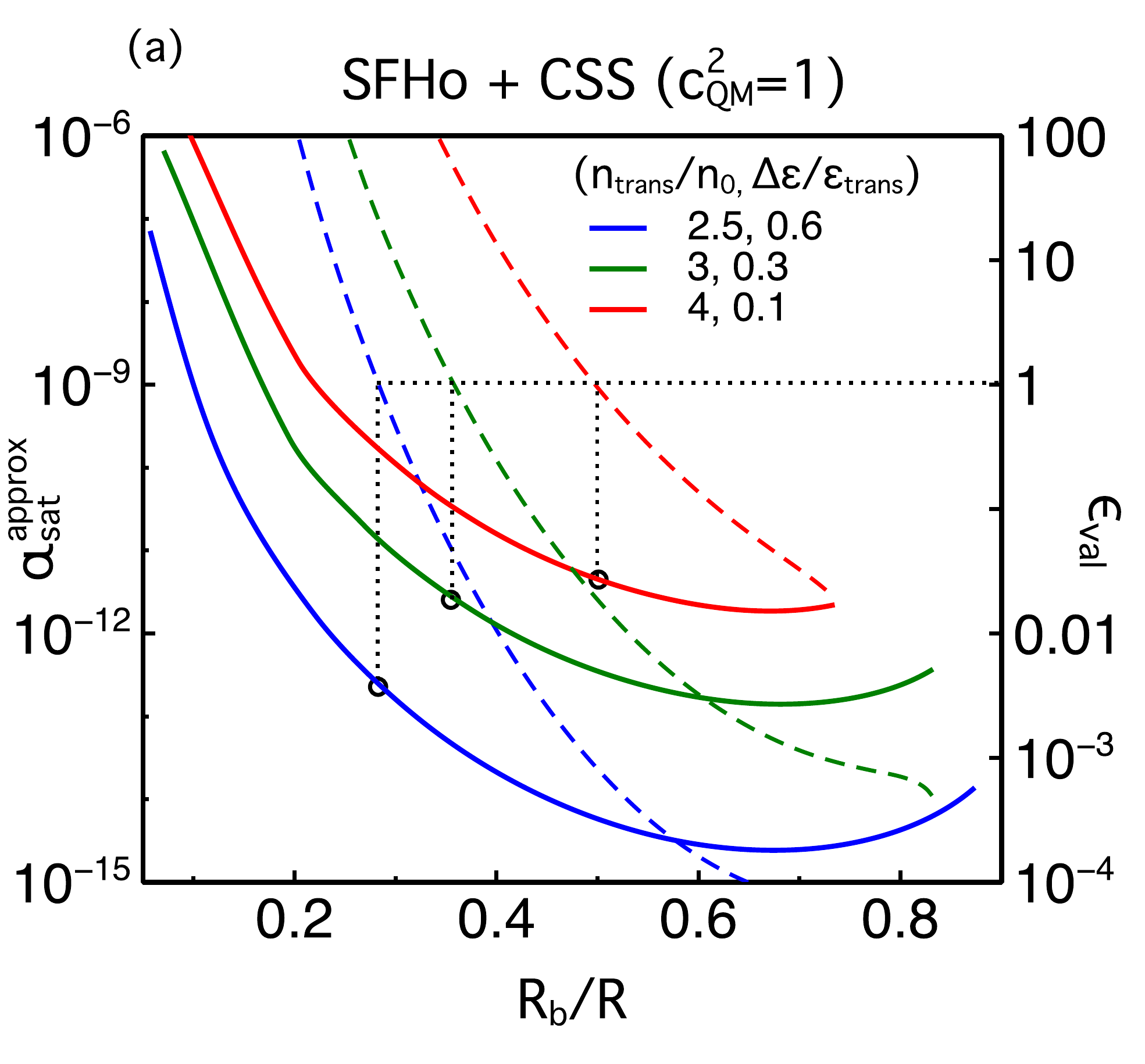}\\[-2ex]
}
\parbox{0.534\hsize}{
\includegraphics[width=\hsize]{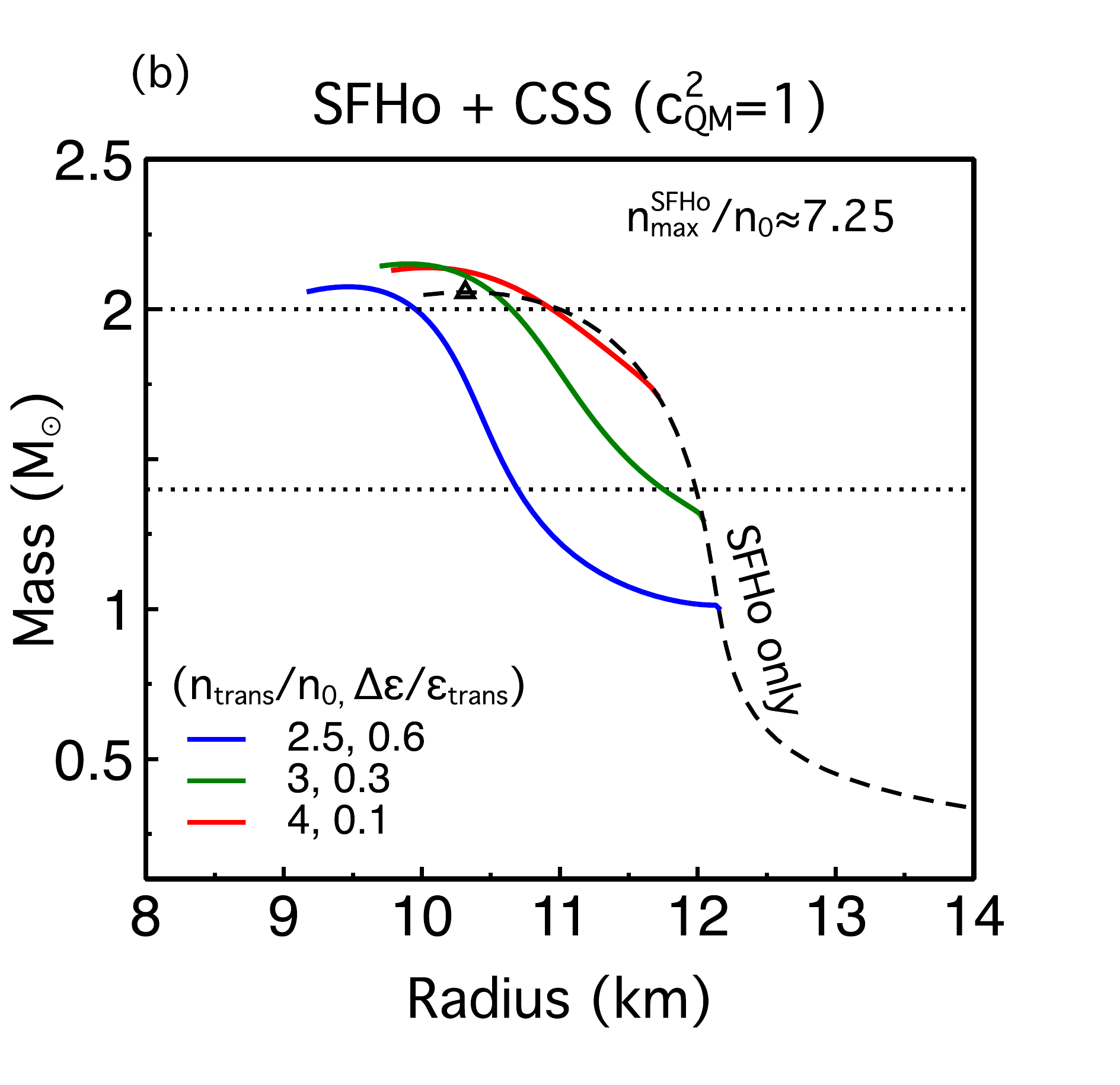}\\[-2ex]
}
\caption{Sensitivity of the r-mode saturation amplitude to phase transition parameters. Panel (a): the validity parameter values (dashed curves) $\eps_{\rm val}$ (right y-axis) decreases with the core size of hybrid stars for three parameter sets (SFHo + CSS). The approximate values for r-mode saturation amplitude (solid curves) $\al_{\rm sat}^{\rm approx}$ (left y-axis) of phase-conversion dissipation is justified only when $\eps_{\rm val}\lesssim 1$, close to the limit when PCD fails to counteract gravitational radiation. Panel (b): mass-radius diagram for the same hybrid SFHo + CSS EoSs in panel (a); the triangle represents the maximum-mass neutron star based on SFHo only (black dashed), with its central density $n_{\rm max}\approx 7.25\, n_0$.
} 
\label{fig:ph-con-eos}
\end{figure*}

FIGURE \ref{fig:ph-con-alpha}(b) displays the dissipated power as a function of the r-mode amplitude for a hybrid star rotating with frequency $f=600\,{\rm Hz}$, with its quark core size $R_{\rm b}/R=0.56$ and the internal temperature $T=10^{8} K$. The hadronic matter EoS is taken from \cite{Hebeler:2010jx}; for the quark matter EoS we use CSS parameters $\ntrans=4.0 \,n_0$, $\De\ep/\etrans=0.2$, and $\cQMsq=1$. Note that the hybrid branch is connected to the hadronic branch in this case, due to the smallness in the energy density discontinuity of the EoS (a weak sharp transition).

At ultra-low amplitude of r-mode oscillations, the phase-conversion dissipation ($P_{\rm dis}\propto \al^3$, analytical approximation represented by the dashed curve) is suppressed relative to the gravitational radiation ($P_{\rm gr}\propto \al^2$) and therefore plays no role in damping the r-mode. If other damping mechanisms are too weak to suppress the r-mode, its amplitude will grow. However, as the amplitude grows, the phase-conversion dissipation becomes stronger, and in this example there is a saturation amplitude $\al_{\rm sat}$ at which it equals the gravitational radiation, and the mode stops growing. At high amplitude in the suprathermal regime ($T \ll \De\mu \ll\mu$) which corresponds to large density/pressure oscillations not assumed true here, the dissipated power is proportional to $\al^2$.

Varying parameters such as the size of the quark core, the rotation frequency, or temperature of the star will shift these curves, and if PCD is too weak then there will be no intersection point ($P_{\rm gr}$ will be greater than $P_{\rm dis}$ at all $\al$) and the growth of the mode cannot be stopped by PCD. However, we can see that if saturation occurs, the resultant $\al_{\rm sat}$ would be in the low-amplitude regime, where an analytical approach is available ($P_{\rm dis}\approx P_{\rm dis}^{\rm sub}$) and the saturation amplitude is extraordinarily low, of order $10^{-12}$. 

FIGURE \ref{fig:ph-con-eos} shows the results when varying phase transition parameters, while the hadronic matter is described by the widely-used SFHo EoS~\cite{Steiner:2012rk}. On panel (a) two quantities are plotted against the quark core ratio $R_{\rm b}/R$ for each hybrid EoS, the validity parameter $\eps_{\rm val}$ and the approximate saturation amplitude $\al_{\rm sat}^{\rm approx}$. Panel (b) displays the mass-radius diagram for the hybrid EoSs applied. $\al_{\rm sat}^{\rm approx}$ is obtained by setting $P_{\rm dis}^{\rm sub} = P_{\rm gr}$ as shown in FIGURE \ref{fig:ph-con-alpha}(b), and to leading order it is a good approximation to the true value of $\al_{\rm sat}$ calculated numerically for $P_{\rm dis} = P_{\rm gr}$. Requiring the next-to-leading (NLO) contribution to be less than a fraction of the total dissipated power yields the validity condition $\eps_{\rm val}\lesssim 1$. It is also found that the mass or quark core size associated with such condition is close to the limiting case where PCD will not stop the growth of r-mode. As can be seen from the figure, this minimum mass or minimum quark core size is sensitive to the phase transition density $\ntrans$: for $\ntrans=2.5 \, n_0$, most of the hybrid stars $\gtrsim 1\,\Msolar$ with $R_{\rm b}/R \gtrsim 0.3$ can have effective damping, whereas for $\ntrans=4.0 \, n_0$, only massive stars $\approx 2 \, \Msolar$ with $R_{\rm b}/R \gtrsim 0.5$ qualify. It will be interesting to test if this trend still holds for hybrid stars on the ``disconnected'' branch (a strong sharp transition), which normally entail larger $\De\ep\gtrsim 0.7$ and lower transition densities $\ntrans\lesssim 2.5 \, n_0$ to satisfy the $2\,\Msolar$ constraint when a soft EoS such as SFHo is chosen, implying all neutron stars we observe are in fact hybrid stars.

\section{SUMMARY AND OUTLOOK}
\label{sec:con} 

Gravitational wave-based probes of neutron star structure have smaller systematic uncertainties than electromagnetic probes, and it is likely that GW detectors may eventually be better at measuring radii of NSs than X-ray satellites. In the near future, with advanced LIGO upgraded and several newly-built detectors become online, further observations of multiple merger events are expected to impose effective constraints on the NS internal composition. Possible precise measurements of the  moment of inertia via pulsar timing~\cite{Lattimer:2004nj,Watts:2014tja} in the next decade, which combined with testing the universal relation, hold promise for distinguishing among EoS models with and without strong phase transitions. Moreover, plenty of work are being carried out on the side of nuclear theory and experiment, aiming to narrow down uncertainties in the pure neutron matter EoS around 1-2 times saturation density, and derive constraints from symmetric nuclear matter at even higher densities~\cite{Danielewicz:2002pu}; on the other hand, appropriate incorporation of phase transitions model-independently is still lacking in the modeling of data and/or numerical simulations.\\

\section{Acknowledgement}
The author would like to thank members of the N3AS (the Network for Neutrinos, Nuclear Astrophysics, and Symmetries) collaboration for helpful discussions, the anonymous referee for instructive comments, and the workshop organizers for their great efforts. This work was supported by the National Science Foundation, Grant PHY-1630782, and the Heising-Simons Foundation, Grant 2017-228.

\bibliographystyle{aipnum-cp}
\bigskip
\bibliography{pt_AIP}{}

\begin{thebibliography}{42}%
\makeatletter
\providecommand \@ifxundefined [1]{%
 \@ifx{#1\undefined}
}%
\providecommand \@ifnum [1]{%
 \ifnum #1\expandafter \@firstoftwo
 \else \expandafter \@secondoftwo
 \fi
}%
\providecommand \@ifx [1]{%
 \ifx #1\expandafter \@firstoftwo
 \else \expandafter \@secondoftwo
 \fi
}%
\providecommand \natexlab [1]{#1}%
\providecommand \enquote  [1]{``#1''}%
\providecommand \bibnamefont  [1]{#1}%
\providecommand \bibfnamefont [1]{#1}%
\providecommand \citenamefont [1]{#1}%
\providecommand \href@noop [0]{\@secondoftwo}%
\providecommand \href [0]{\begingroup \@sanitize@url \@href}%
\providecommand \@href[1]{\@@startlink{#1}\@@href}%
\providecommand \@@href[1]{\endgroup#1\@@endlink}%
\providecommand \@sanitize@url [0]{\catcode `\$12\catcode `\&12\catcode
  `\#12\catcode `\^12\catcode `\_12\catcode `\%12\relax}%
\providecommand \@@startlink[1]{}%
\providecommand \@@endlink[0]{}%
\providecommand \url  [0]{\begingroup\@sanitize@url \@url }%
\providecommand \@url [1]{\endgroup\@href {#1}{\urlprefix }}%
\providecommand \urlprefix  [0]{URL }%
\providecommand \Eprint [0]{\href }%
\providecommand \doibase [0]{http://dx.doi.org/}%
\providecommand \selectlanguage [0]{\@gobble}%
\providecommand \bibinfo  [0]{\@secondoftwo}%
\providecommand \bibfield  [0]{\@secondoftwo}%
\providecommand \translation [1]{[#1]}%
\providecommand \BibitemOpen [0]{}%
\providecommand \bibitemStop [0]{}%
\providecommand \bibitemNoStop [0]{.\EOS\space}%
\providecommand \EOS [0]{\spacefactor3000\relax}%
\providecommand \BibitemShut  [1]{\csname bibitem#1\endcsname}%
\let\auto@bib@innerbib\@empty
\bibitem [{\citenamefont {Demorest}\ \emph {et~al.}(2010)\citenamefont
  {Demorest}, \citenamefont {Pennucci}, \citenamefont {Ransom}, \citenamefont
  {Roberts},\ and\ \citenamefont {Hessels}}]{Demorest:2010bx}%
  \BibitemOpen
  \bibfield  {author} {\bibinfo {author} {\bibfnamefont {P.}~\bibnamefont
  {Demorest}}, \bibinfo {author} {\bibfnamefont {T.}~\bibnamefont {Pennucci}},
  \bibinfo {author} {\bibfnamefont {S.}~\bibnamefont {Ransom}}, \bibinfo
  {author} {\bibfnamefont {M.}~\bibnamefont {Roberts}}, \ and\ \bibinfo
  {author} {\bibfnamefont {J.}~\bibnamefont {Hessels}},\ }\href {\doibase
  10.1038/nature09466} {\bibfield  {journal} {\bibinfo  {journal} {Nature}\
  }\textbf {\bibinfo {volume} {467}},\ \unskip\ \bibinfo {pages} {1081--1083}
  (\bibinfo {year} {2010})},\ \Eprint {http://arxiv.org/abs/1010.5788}
  {arXiv:1010.5788 [astro-ph.HE]} \BibitemShut {NoStop}%
\bibitem [{\citenamefont {Antoniadis}\ \emph {et~al.}(2013)\citenamefont
  {Antoniadis}, \citenamefont {Freire}, \citenamefont {Wex}, \citenamefont
  {Tauris}, \citenamefont {Lynch} \emph {et~al.}}]{Antoniadis:2013pzd}%
  \BibitemOpen
  \bibfield  {author} {\bibinfo {author} {\bibfnamefont {J.}~\bibnamefont
  {Antoniadis}}, \bibinfo {author} {\bibfnamefont {P.~C.}\ \bibnamefont
  {Freire}}, \bibinfo {author} {\bibfnamefont {N.}~\bibnamefont {Wex}},
  \bibinfo {author} {\bibfnamefont {T.~M.}\ \bibnamefont {Tauris}}, \bibinfo
  {author} {\bibfnamefont {R.~S.}\ \bibnamefont {Lynch}},  \emph {et~al.},\
  }\href {\doibase 10.1126/science.1233232} {\bibfield  {journal} {\bibinfo
  {journal} {Science}\ }\textbf {\bibinfo {volume} {340}},\ p.\ \bibinfo
  {pages} {6131} (\bibinfo {year} {2013})},\ \Eprint
  {http://arxiv.org/abs/1304.6875} {arXiv:1304.6875 [astro-ph.HE]} \BibitemShut
  {NoStop}%
\bibitem [{\citenamefont {Abbott}\ \emph {et~al.}(2017)\citenamefont {Abbott}
  \emph {et~al.}}]{LIGO:2017qsa}%
  \BibitemOpen
  \bibfield  {author} {\bibinfo {author} {\bibfnamefont {B.~P.}\ \bibnamefont
  {Abbott}} \emph {et~al.} (\bibinfo {collaboration} {Virgo, LIGO
  Scientific}),\ }\href {\doibase 10.1103/PhysRevLett.119.161101} {\bibfield
  {journal} {\bibinfo  {journal} {Phys. Rev. Lett.}\ }\textbf {\bibinfo
  {volume} {119}},\ p.\ \bibinfo {pages} {161101} (\bibinfo {year} {2017})},\
  \Eprint {http://arxiv.org/abs/1710.05832} {arXiv:1710.05832 [gr-qc]}
  \BibitemShut {NoStop}%
\bibitem [{\citenamefont {Tolman}(1939)}]{Tolman:1939jz}%
  \BibitemOpen
  \bibfield  {author} {\bibinfo {author} {\bibfnamefont {R.~C.}\ \bibnamefont
  {Tolman}},\ }\href {\doibase 10.1103/PhysRev.55.364} {\bibfield  {journal}
  {\bibinfo  {journal} {Phys. Rev.}\ }\textbf {\bibinfo {volume} {55}},\
  \unskip\ \bibinfo {pages} {364--373} (\bibinfo {year} {1939})}\BibitemShut
  {NoStop}%
\bibitem [{\citenamefont {Oppenheimer}\ and\ \citenamefont
  {Volkoff}(1939)}]{Oppenheimer:1939ne}%
  \BibitemOpen
  \bibfield  {author} {\bibinfo {author} {\bibfnamefont {J.~R.}\ \bibnamefont
  {Oppenheimer}}\ and\ \bibinfo {author} {\bibfnamefont {G.~M.}\ \bibnamefont
  {Volkoff}},\ }\href {\doibase 10.1103/PhysRev.55.374} {\bibfield  {journal}
  {\bibinfo  {journal} {Phys. Rev.}\ }\textbf {\bibinfo {volume} {55}},\
  \unskip\ \bibinfo {pages} {374--381} (\bibinfo {year} {1939})}\BibitemShut
  {NoStop}%
\bibitem [{\citenamefont {Hebeler}\ \emph {et~al.}(2010)\citenamefont
  {Hebeler}, \citenamefont {Lattimer}, \citenamefont {Pethick},\ and\
  \citenamefont {Schwenk}}]{Hebeler:2010jx}%
  \BibitemOpen
  \bibfield  {author} {\bibinfo {author} {\bibfnamefont {K.}~\bibnamefont
  {Hebeler}}, \bibinfo {author} {\bibfnamefont {J.~M.}\ \bibnamefont
  {Lattimer}}, \bibinfo {author} {\bibfnamefont {C.~J.}\ \bibnamefont
  {Pethick}}, \ and\ \bibinfo {author} {\bibfnamefont {A.}~\bibnamefont
  {Schwenk}},\ }\href {\doibase 10.1103/PhysRevLett.105.161102} {\bibfield
  {journal} {\bibinfo  {journal} {Phys. Rev. Lett.}\ }\textbf {\bibinfo
  {volume} {105}},\ p.\ \bibinfo {pages} {161102} (\bibinfo {year} {2010})},\
  \Eprint {http://arxiv.org/abs/1007.1746} {arXiv:1007.1746 [nucl-th]}
  \BibitemShut {NoStop}%
\bibitem [{\citenamefont {Gandolfi}, \citenamefont {Carlson},\ and\
  \citenamefont {Reddy}(2012)}]{Gandolfi:2011xu}%
  \BibitemOpen
  \bibfield  {author} {\bibinfo {author} {\bibfnamefont {S.}~\bibnamefont
  {Gandolfi}}, \bibinfo {author} {\bibfnamefont {J.}~\bibnamefont {Carlson}}, \
  and\ \bibinfo {author} {\bibfnamefont {S.}~\bibnamefont {Reddy}},\ }\href
  {\doibase 10.1103/PhysRevC.85.032801} {\bibfield  {journal} {\bibinfo
  {journal} {Phys. Rev. C}\ }\textbf {\bibinfo {volume} {85}},\ p.\ \bibinfo
  {pages} {032801} (\bibinfo {year} {2012})},\ \Eprint
  {http://arxiv.org/abs/1101.1921} {arXiv:1101.1921 [nucl-th]} \BibitemShut
  {NoStop}%
\bibitem [{\citenamefont {Masuda}, \citenamefont {Hatsuda},\ and\ \citenamefont
  {Takatsuka}(2013)}]{Masuda:2012ed}%
  \BibitemOpen
  \bibfield  {author} {\bibinfo {author} {\bibfnamefont {K.}~\bibnamefont
  {Masuda}}, \bibinfo {author} {\bibfnamefont {T.}~\bibnamefont {Hatsuda}}, \
  and\ \bibinfo {author} {\bibfnamefont {T.}~\bibnamefont {Takatsuka}},\ }\href
  {\doibase 10.1093/ptep/ptt045} {\bibfield  {journal} {\bibinfo  {journal}
  {PTEP}\ }\textbf {\bibinfo {volume} {2013}},\ p.\ \bibinfo {pages} {073D01}
  (\bibinfo {year} {2013})},\ \Eprint {http://arxiv.org/abs/1212.6803}
  {arXiv:1212.6803 [nucl-th]} \BibitemShut {NoStop}%
\bibitem [{\citenamefont {Fukushima}\ and\ \citenamefont
  {Kojo}(2016)}]{Fukushima:2015bda}%
  \BibitemOpen
  \bibfield  {author} {\bibinfo {author} {\bibfnamefont {K.}~\bibnamefont
  {Fukushima}}\ and\ \bibinfo {author} {\bibfnamefont {T.}~\bibnamefont
  {Kojo}},\ }\href {\doibase 10.3847/0004-637X/817/2/180} {\bibfield  {journal}
  {\bibinfo  {journal} {Astrophys. J.}\ }\textbf {\bibinfo {volume} {817}},\
  p.\ \bibinfo {pages} {180} (\bibinfo {year} {2016})},\ \Eprint
  {http://arxiv.org/abs/1509.00356} {arXiv:1509.00356 [nucl-th]} \BibitemShut
  {NoStop}%
\bibitem [{\citenamefont {Baym}\ \emph {et~al.}(2018)\citenamefont {Baym},
  \citenamefont {Hatsuda}, \citenamefont {Kojo}, \citenamefont {Powell},
  \citenamefont {Song},\ and\ \citenamefont {Takatsuka}}]{Baym:2017whm}%
  \BibitemOpen
  \bibfield  {author} {\bibinfo {author} {\bibfnamefont {G.}~\bibnamefont
  {Baym}}, \bibinfo {author} {\bibfnamefont {T.}~\bibnamefont {Hatsuda}},
  \bibinfo {author} {\bibfnamefont {T.}~\bibnamefont {Kojo}}, \bibinfo {author}
  {\bibfnamefont {P.~D.}\ \bibnamefont {Powell}}, \bibinfo {author}
  {\bibfnamefont {Y.}~\bibnamefont {Song}}, \ and\ \bibinfo {author}
  {\bibfnamefont {T.}~\bibnamefont {Takatsuka}},\ }\href {\doibase
  10.1088/1361-6633/aaae14} {\bibfield  {journal} {\bibinfo  {journal} {Rept.
  Prog. Phys.}\ }\textbf {\bibinfo {volume} {81}},\ p.\ \bibinfo {pages}
  {056902} (\bibinfo {year} {2018})},\ \Eprint
  {http://arxiv.org/abs/1707.04966} {arXiv:1707.04966 [astro-ph.HE]}
  \BibitemShut {NoStop}%
\bibitem [{\citenamefont {McLerran}\ and\ \citenamefont
  {Reddy}(2019)}]{McLerran:2018hbz}%
  \BibitemOpen
  \bibfield  {author} {\bibinfo {author} {\bibfnamefont {L.}~\bibnamefont
  {McLerran}}\ and\ \bibinfo {author} {\bibfnamefont {S.}~\bibnamefont
  {Reddy}},\ }\href {\doibase 10.1103/PhysRevLett.122.122701} {\bibfield
  {journal} {\bibinfo  {journal} {Phys. Rev. Lett.}\ }\textbf {\bibinfo
  {volume} {122}},\ p.\ \bibinfo {pages} {122701} (\bibinfo {year} {2019})},\
  \Eprint {http://arxiv.org/abs/1811.12503} {arXiv:1811.12503 [nucl-th]}
  \BibitemShut {NoStop}%
\bibitem [{\citenamefont {Borsanyi}\ \emph {et~al.}(2012)\citenamefont
  {Borsanyi}, \citenamefont {Endrodi}, \citenamefont {Fodor}, \citenamefont
  {Katz}, \citenamefont {Krieg}, \citenamefont {Ratti},\ and\ \citenamefont
  {Szabo}}]{Borsanyi:2012cr}%
  \BibitemOpen
  \bibfield  {author} {\bibinfo {author} {\bibfnamefont {S.}~\bibnamefont
  {Borsanyi}}, \bibinfo {author} {\bibfnamefont {G.}~\bibnamefont {Endrodi}},
  \bibinfo {author} {\bibfnamefont {Z.}~\bibnamefont {Fodor}}, \bibinfo
  {author} {\bibfnamefont {S.~D.}\ \bibnamefont {Katz}}, \bibinfo {author}
  {\bibfnamefont {S.}~\bibnamefont {Krieg}}, \bibinfo {author} {\bibfnamefont
  {C.}~\bibnamefont {Ratti}}, \ and\ \bibinfo {author} {\bibfnamefont {K.~K.}\
  \bibnamefont {Szabo}},\ }\href {\doibase 10.1007/JHEP08(2012)053} {\bibfield
  {journal} {\bibinfo  {journal} {JHEP}\ }\textbf {\bibinfo {volume} {08}},\
  p.\ \bibinfo {pages} {053} (\bibinfo {year} {2012})},\ \Eprint
  {http://arxiv.org/abs/1204.6710} {arXiv:1204.6710 [hep-lat]} \BibitemShut
  {NoStop}%
\bibitem [{\citenamefont {Kurkela}\ \emph {et~al.}(2014)\citenamefont
  {Kurkela}, \citenamefont {Fraga}, \citenamefont {Schaffner-Bielich},\ and\
  \citenamefont {Vuorinen}}]{Kurkela:2014vha}%
  \BibitemOpen
  \bibfield  {author} {\bibinfo {author} {\bibfnamefont {A.}~\bibnamefont
  {Kurkela}}, \bibinfo {author} {\bibfnamefont {E.~S.}\ \bibnamefont {Fraga}},
  \bibinfo {author} {\bibfnamefont {J.}~\bibnamefont {Schaffner-Bielich}}, \
  and\ \bibinfo {author} {\bibfnamefont {A.}~\bibnamefont {Vuorinen}},\ }\href
  {\doibase 10.1088/0004-637X/789/2/127} {\bibfield  {journal} {\bibinfo
  {journal} {Astrophys. J.}\ }\textbf {\bibinfo {volume} {789}},\ p.\ \bibinfo
  {pages} {127} (\bibinfo {year} {2014})},\ \Eprint
  {http://arxiv.org/abs/1402.6618} {arXiv:1402.6618 [astro-ph.HE]} \BibitemShut
  {NoStop}%
\bibitem [{\citenamefont {Bedaque}\ and\ \citenamefont
  {Steiner}(2015)}]{Bedaque:2014sqa}%
  \BibitemOpen
  \bibfield  {author} {\bibinfo {author} {\bibfnamefont {P.~F.}\ \bibnamefont
  {Bedaque}}\ and\ \bibinfo {author} {\bibfnamefont {A.~W.}\ \bibnamefont
  {Steiner}},\ }\href {\doibase 10.1103/PhysRevLett.114.031103} {\bibfield
  {journal} {\bibinfo  {journal} {Phys. Rev. Lett.}\ }\textbf {\bibinfo
  {volume} {114}},\ p.\ \bibinfo {pages} {031103} (\bibinfo {year} {2015})},\
  \Eprint {http://arxiv.org/abs/1408.5116} {arXiv:1408.5116 [nucl-th]}
  \BibitemShut {NoStop}%
\bibitem [{\citenamefont {Alford}, \citenamefont {Han},\ and\ \citenamefont
  {Prakash}(2013)}]{Alford:2013aca}%
  \BibitemOpen
  \bibfield  {author} {\bibinfo {author} {\bibfnamefont {M.~G.}\ \bibnamefont
  {Alford}}, \bibinfo {author} {\bibfnamefont {S.}~\bibnamefont {Han}}, \ and\
  \bibinfo {author} {\bibfnamefont {M.}~\bibnamefont {Prakash}},\ }\href
  {\doibase 10.1103/PhysRevD.88.083013} {\bibfield  {journal} {\bibinfo
  {journal} {Phys. Rev. D}\ }\textbf {\bibinfo {volume} {88}},\ p.\ \bibinfo
  {pages} {083013} (\bibinfo {year} {2013})},\ \Eprint
  {http://arxiv.org/abs/1302.4732} {arXiv:1302.4732 [astro-ph.SR]} \BibitemShut
  {NoStop}%
\bibitem [{\citenamefont {Alford}\ and\ \citenamefont
  {Han}(2016)}]{Alford:2015gna}%
  \BibitemOpen
  \bibfield  {author} {\bibinfo {author} {\bibfnamefont {M.~G.}\ \bibnamefont
  {Alford}}\ and\ \bibinfo {author} {\bibfnamefont {S.}~\bibnamefont {Han}},\
  }\href {\doibase 10.1140/epja/i2016-16062-9} {\bibfield  {journal} {\bibinfo
  {journal} {Eur. Phys. J.}\ }\textbf {\bibinfo {volume} {A52}},\ p.~\bibinfo
  {pages} {62} (\bibinfo {year} {2016})},\ \Eprint
  {http://arxiv.org/abs/1508.01261} {arXiv:1508.01261 [nucl-th]} \BibitemShut
  {NoStop}%
\bibitem [{\citenamefont {Wade}\ \emph {et~al.}(2014)\citenamefont {Wade},
  \citenamefont {Creighton}, \citenamefont {Ochsner}, \citenamefont {Lackey},
  \citenamefont {Farr}, \citenamefont {Littenberg},\ and\ \citenamefont
  {Raymond}}]{Wade:2014vqa}%
  \BibitemOpen
  \bibfield  {author} {\bibinfo {author} {\bibfnamefont {L.}~\bibnamefont
  {Wade}}, \bibinfo {author} {\bibfnamefont {J.~D.~E.}\ \bibnamefont
  {Creighton}}, \bibinfo {author} {\bibfnamefont {E.}~\bibnamefont {Ochsner}},
  \bibinfo {author} {\bibfnamefont {B.~D.}\ \bibnamefont {Lackey}}, \bibinfo
  {author} {\bibfnamefont {B.~F.}\ \bibnamefont {Farr}}, \bibinfo {author}
  {\bibfnamefont {T.~B.}\ \bibnamefont {Littenberg}}, \ and\ \bibinfo {author}
  {\bibfnamefont {V.}~\bibnamefont {Raymond}},\ }\href {\doibase
  10.1103/PhysRevD.89.103012} {\bibfield  {journal} {\bibinfo  {journal} {Phys.
  Rev. D}\ }\textbf {\bibinfo {volume} {89}},\ p.\ \bibinfo {pages} {103012}
  (\bibinfo {year} {2014})},\ \Eprint {http://arxiv.org/abs/1402.5156}
  {arXiv:1402.5156 [gr-qc]} \BibitemShut {NoStop}%
\bibitem [{\citenamefont {Gross-Boelting}, \citenamefont {Fuchs},\ and\
  \citenamefont {Faessler}(1999)}]{GrossBoelting:1998jg}%
  \BibitemOpen
  \bibfield  {author} {\bibinfo {author} {\bibfnamefont {T.}~\bibnamefont
  {Gross-Boelting}}, \bibinfo {author} {\bibfnamefont {C.}~\bibnamefont
  {Fuchs}}, \ and\ \bibinfo {author} {\bibfnamefont {A.}~\bibnamefont
  {Faessler}},\ }\href {\doibase 10.1016/S0375-9474(99)00022-6} {\bibfield
  {journal} {\bibinfo  {journal} {Nucl. Phys.}\ }\textbf {\bibinfo {volume}
  {A648}},\ \unskip\ \bibinfo {pages} {105--137} (\bibinfo {year} {1999})},\
  \Eprint {http://arxiv.org/abs/nucl-th/9810071} {arXiv:nucl-th/9810071
  [nucl-th]} \BibitemShut {NoStop}%
\bibitem [{\citenamefont {Cromartie}\ \emph {et~al.}(2019)\citenamefont
  {Cromartie} \emph {et~al.}}]{Cromartie:2019kug}%
  \BibitemOpen
  \bibfield  {author} {\bibinfo {author} {\bibfnamefont {H.~T.}\ \bibnamefont
  {Cromartie}} \emph {et~al.},\ }\href@noop {} {\bibfield  {journal} {\bibinfo
  {journal} {arXiv e-prints}\ } (\bibinfo {year} {2019})},\ \Eprint
  {http://arxiv.org/abs/1904.06759} {arXiv:1904.06759 [astro-ph.HE]}
  \BibitemShut {NoStop}%
\bibitem [{\citenamefont {Han}\ and\ \citenamefont
  {Steiner}(2018)}]{Han:2018mtj}%
  \BibitemOpen
  \bibfield  {author} {\bibinfo {author} {\bibfnamefont {S.}~\bibnamefont
  {Han}}\ and\ \bibinfo {author} {\bibfnamefont {A.~W.}\ \bibnamefont
  {Steiner}},\ }\href@noop {} {\bibfield  {journal} {\bibinfo  {journal} {arXiv
  e-prints}\ } (\bibinfo {year} {2018})},\ \Eprint
  {http://arxiv.org/abs/1810.10967} {arXiv:1810.10967 [nucl-th]} \BibitemShut
  {NoStop}%
\bibitem [{\citenamefont {Alford}\ and\ \citenamefont
  {Sedrakian}(2017)}]{Alford:2017qgh}%
  \BibitemOpen
  \bibfield  {author} {\bibinfo {author} {\bibfnamefont {M.~G.}\ \bibnamefont
  {Alford}}\ and\ \bibinfo {author} {\bibfnamefont {A.}~\bibnamefont
  {Sedrakian}},\ }\href {\doibase 10.1103/PhysRevLett.119.161104} {\bibfield
  {journal} {\bibinfo  {journal} {Phys. Rev. Lett.}\ }\textbf {\bibinfo
  {volume} {119}},\ p.\ \bibinfo {pages} {161104} (\bibinfo {year} {2017})},\
  \Eprint {http://arxiv.org/abs/1706.01592} {arXiv:1706.01592 [astro-ph.HE]}
  \BibitemShut {NoStop}%
\bibitem [{\citenamefont {Abbott}\ \emph {et~al.}(2019)\citenamefont {Abbott}
  \emph {et~al.}}]{LIGO:2018wiz}%
  \BibitemOpen
  \bibfield  {author} {\bibinfo {author} {\bibfnamefont {B.~P.}\ \bibnamefont
  {Abbott}} \emph {et~al.} (\bibinfo {collaboration} {LIGO Scientific,
  Virgo}),\ }\href {\doibase 10.1103/PhysRevX.9.011001} {\bibfield  {journal}
  {\bibinfo  {journal} {Phys. Rev. X}\ }\textbf {\bibinfo {volume} {9}},\ p.\
  \bibinfo {pages} {011001} (\bibinfo {year} {2019})},\ \Eprint
  {http://arxiv.org/abs/1805.11579} {arXiv:1805.11579 [gr-qc]} \BibitemShut
  {NoStop}%
\bibitem [{\citenamefont {Yagi}\ and\ \citenamefont
  {Yunes}(2013)}]{Yagi:2013awa}%
  \BibitemOpen
  \bibfield  {author} {\bibinfo {author} {\bibfnamefont {K.}~\bibnamefont
  {Yagi}}\ and\ \bibinfo {author} {\bibfnamefont {N.}~\bibnamefont {Yunes}},\
  }\href {\doibase 10.1103/PhysRevD.88.023009} {\bibfield  {journal} {\bibinfo
  {journal} {Phys. Rev. D}\ }\textbf {\bibinfo {volume} {88}},\ p.\ \bibinfo
  {pages} {023009} (\bibinfo {year} {2013})},\ \Eprint
  {http://arxiv.org/abs/1303.1528} {arXiv:1303.1528 [gr-qc]} \BibitemShut
  {NoStop}%
\bibitem [{\citenamefont {Yagi}\ and\ \citenamefont
  {Yunes}(2017)}]{Yagi:2016bkt}%
  \BibitemOpen
  \bibfield  {author} {\bibinfo {author} {\bibfnamefont {K.}~\bibnamefont
  {Yagi}}\ and\ \bibinfo {author} {\bibfnamefont {N.}~\bibnamefont {Yunes}},\
  }\href {\doibase 10.1016/j.physrep.2017.03.002} {\bibfield  {journal}
  {\bibinfo  {journal} {Phys. Rept.}\ }\textbf {\bibinfo {volume} {681}},\
  \unskip\ \bibinfo {pages} {1--72} (\bibinfo {year} {2017})},\ \Eprint
  {http://arxiv.org/abs/1608.02582} {arXiv:1608.02582 [gr-qc]} \BibitemShut
  {NoStop}%
\bibitem [{\citenamefont {Steiner}\ \emph {et~al.}(2005)\citenamefont
  {Steiner}, \citenamefont {Prakash}, \citenamefont {Lattimer},\ and\
  \citenamefont {Ellis}}]{Steiner:2004fi}%
  \BibitemOpen
  \bibfield  {author} {\bibinfo {author} {\bibfnamefont {A.~W.}\ \bibnamefont
  {Steiner}}, \bibinfo {author} {\bibfnamefont {M.}~\bibnamefont {Prakash}},
  \bibinfo {author} {\bibfnamefont {J.~M.}\ \bibnamefont {Lattimer}}, \ and\
  \bibinfo {author} {\bibfnamefont {P.~J.}\ \bibnamefont {Ellis}},\ }\href
  {\doibase 10.1016/j.physrep.2005.02.004} {\bibfield  {journal} {\bibinfo
  {journal} {Phys. Rept.}\ }\textbf {\bibinfo {volume} {411}},\ \unskip\
  \bibinfo {pages} {325--375} (\bibinfo {year} {2005})},\ \Eprint
  {http://arxiv.org/abs/nucl-th/0410066} {arXiv:nucl-th/0410066 [nucl-th]}
  \BibitemShut {NoStop}%
\bibitem [{\citenamefont {Page}\ \emph {et~al.}(2013)\citenamefont {Page},
  \citenamefont {Lattimer}, \citenamefont {Prakash},\ and\ \citenamefont
  {Steiner}}]{Page:2013hxa}%
  \BibitemOpen
  \bibfield  {author} {\bibinfo {author} {\bibfnamefont {D.}~\bibnamefont
  {Page}}, \bibinfo {author} {\bibfnamefont {J.~M.}\ \bibnamefont {Lattimer}},
  \bibinfo {author} {\bibfnamefont {M.}~\bibnamefont {Prakash}}, \ and\
  \bibinfo {author} {\bibfnamefont {A.~W.}\ \bibnamefont {Steiner}},\
  }\href@noop {} {\  (\bibinfo {year} {2013})},\ \Eprint
  {http://arxiv.org/abs/1302.6626} {arXiv:1302.6626 [astro-ph.HE]} \BibitemShut
  {NoStop}%
\bibitem [{\citenamefont {Brown}\ \emph {et~al.}(2018)\citenamefont {Brown},
  \citenamefont {Cumming}, \citenamefont {Fattoyev}, \citenamefont {Horowitz},
  \citenamefont {Page},\ and\ \citenamefont {Reddy}}]{Brown:2017gxd}%
  \BibitemOpen
  \bibfield  {author} {\bibinfo {author} {\bibfnamefont {E.~F.}\ \bibnamefont
  {Brown}}, \bibinfo {author} {\bibfnamefont {A.}~\bibnamefont {Cumming}},
  \bibinfo {author} {\bibfnamefont {F.~J.}\ \bibnamefont {Fattoyev}}, \bibinfo
  {author} {\bibfnamefont {C.~J.}\ \bibnamefont {Horowitz}}, \bibinfo {author}
  {\bibfnamefont {D.}~\bibnamefont {Page}}, \ and\ \bibinfo {author}
  {\bibfnamefont {S.}~\bibnamefont {Reddy}},\ }\href {\doibase
  10.1103/PhysRevLett.120.182701} {\bibfield  {journal} {\bibinfo  {journal}
  {Phys. Rev. Lett.}\ }\textbf {\bibinfo {volume} {120}},\ p.\ \bibinfo {pages}
  {182701} (\bibinfo {year} {2018})},\ \Eprint
  {http://arxiv.org/abs/1801.00041} {arXiv:1801.00041 [astro-ph.HE]}
  \BibitemShut {NoStop}%
\bibitem [{\citenamefont {Heinke}\ \emph {et~al.}(2010)\citenamefont {Heinke}
  \emph {et~al.}}]{Heinke10}%
  \BibitemOpen
  \bibfield  {author} {\bibinfo {author} {\bibfnamefont {C.~O.}\ \bibnamefont
  {Heinke}} \emph {et~al.},\ }\href {\doibase 10.1088/0004-637X/714/1/894}
  {\bibfield  {journal} {\bibinfo  {journal} {Astrophys. J.}\ }\textbf
  {\bibinfo {volume} {714}},\ \unskip\ \bibinfo {pages} {894--903} (\bibinfo
  {year} {2010})},\ \Eprint {http://arxiv.org/abs/0911.0444} {arXiv:0911.0444
  [astro-ph.HE]} \BibitemShut {NoStop}%
\bibitem [{\citenamefont {Wijnands}, \citenamefont {Degenaar},\ and\
  \citenamefont {Page}(2017)}]{Wijnands:2017jsc}%
  \BibitemOpen
  \bibfield  {author} {\bibinfo {author} {\bibfnamefont {R.}~\bibnamefont
  {Wijnands}}, \bibinfo {author} {\bibfnamefont {N.}~\bibnamefont {Degenaar}},
  \ and\ \bibinfo {author} {\bibfnamefont {D.}~\bibnamefont {Page}},\ }\href
  {\doibase 10.1007/s12036-017-9466-5} {\bibfield  {journal} {\bibinfo
  {journal} {J. Astrophys. Astron.}\ }\textbf {\bibinfo {volume} {38}},\
  p.~\bibinfo {pages} {49} (\bibinfo {year} {2017})},\ \Eprint
  {http://arxiv.org/abs/1709.07034} {arXiv:1709.07034 [astro-ph.HE]}
  \BibitemShut {NoStop}%
\bibitem [{\citenamefont {Morsink}\ and\ \citenamefont
  {Leahy}(2011)}]{Morsink:2009wv}%
  \BibitemOpen
  \bibfield  {author} {\bibinfo {author} {\bibfnamefont {S.~M.}\ \bibnamefont
  {Morsink}}\ and\ \bibinfo {author} {\bibfnamefont {D.~A.}\ \bibnamefont
  {Leahy}},\ }\href {\doibase 10.1088/0004-637X/726/1/56} {\bibfield  {journal}
  {\bibinfo  {journal} {Astrophys. J.}\ }\textbf {\bibinfo {volume} {726}},\
  p.~\bibinfo {pages} {56} (\bibinfo {year} {2011})},\ \Eprint
  {http://arxiv.org/abs/0911.0887} {arXiv:0911.0887 [astro-ph.HE]} \BibitemShut
  {NoStop}%
\bibitem [{\citenamefont {Han}\ and\ \citenamefont
  {Steiner}(2017)}]{Han:2017jaj}%
  \BibitemOpen
  \bibfield  {author} {\bibinfo {author} {\bibfnamefont {S.}~\bibnamefont
  {Han}}\ and\ \bibinfo {author} {\bibfnamefont {A.~W.}\ \bibnamefont
  {Steiner}},\ }\href {\doibase 10.1103/PhysRevC.96.035802} {\bibfield
  {journal} {\bibinfo  {journal} {Phys. Rev. C}\ }\textbf {\bibinfo {volume}
  {96}},\ p.\ \bibinfo {pages} {035802} (\bibinfo {year} {2017})},\ \Eprint
  {http://arxiv.org/abs/1702.08452} {arXiv:1702.08452 [astro-ph.HE]}
  \BibitemShut {NoStop}%
\bibitem [{\citenamefont {Andersson}\ and\ \citenamefont
  {Kokkotas}(2001)}]{Andersson:2000mf}%
  \BibitemOpen
  \bibfield  {author} {\bibinfo {author} {\bibfnamefont {N.}~\bibnamefont
  {Andersson}}\ and\ \bibinfo {author} {\bibfnamefont {K.~D.}\ \bibnamefont
  {Kokkotas}},\ }\href {\doibase 10.1142/S0218271801001062} {\bibfield
  {journal} {\bibinfo  {journal} {Int. J. Mod. Phys.}\ }\textbf {\bibinfo
  {volume} {D10}},\ \unskip\ \bibinfo {pages} {381--442} (\bibinfo {year}
  {2001})},\ \Eprint {http://arxiv.org/abs/gr-qc/0010102} {arXiv:gr-qc/0010102
  [gr-qc]} \BibitemShut {NoStop}%
\bibitem [{\citenamefont {Haskell}(2015)}]{Haskell:2015iia}%
  \BibitemOpen
  \bibfield  {author} {\bibinfo {author} {\bibfnamefont {B.}~\bibnamefont
  {Haskell}},\ }\href {\doibase 10.1142/S0218301315410074} {\bibfield
  {journal} {\bibinfo  {journal} {Int. J. Mod. Phys.}\ }\textbf {\bibinfo
  {volume} {E24}},\ p.\ \bibinfo {pages} {1541007} (\bibinfo {year} {2015})},\
  \Eprint {http://arxiv.org/abs/1509.04370} {arXiv:1509.04370 [astro-ph.HE]}
  \BibitemShut {NoStop}%
\bibitem [{\citenamefont {Kokkotas}\ and\ \citenamefont
  {Schwenzer}(2016)}]{Kokkotas:2015gea}%
  \BibitemOpen
  \bibfield  {author} {\bibinfo {author} {\bibfnamefont {K.~D.}\ \bibnamefont
  {Kokkotas}}\ and\ \bibinfo {author} {\bibfnamefont {K.}~\bibnamefont
  {Schwenzer}},\ }\href {\doibase 10.1140/epja/i2016-16038-9} {\bibfield
  {journal} {\bibinfo  {journal} {Eur. Phys. J.}\ }\textbf {\bibinfo {volume}
  {A52}},\ p.~\bibinfo {pages} {38} (\bibinfo {year} {2016})},\ \Eprint
  {http://arxiv.org/abs/1510.07051} {arXiv:1510.07051 [gr-qc]} \BibitemShut
  {NoStop}%
\bibitem [{\citenamefont {Mahmoodifar}\ and\ \citenamefont
  {Strohmayer}(2013)}]{Mahmoodifar:2013quw}%
  \BibitemOpen
  \bibfield  {author} {\bibinfo {author} {\bibfnamefont {S.}~\bibnamefont
  {Mahmoodifar}}\ and\ \bibinfo {author} {\bibfnamefont {T.}~\bibnamefont
  {Strohmayer}},\ }\href@noop {} {\bibfield  {journal} {\bibinfo  {journal}
  {Astrophys. J.}\ }\textbf {\bibinfo {volume} {773}},\ p.\ \bibinfo {pages}
  {140} (\bibinfo {year} {2013})},\ \Eprint {http://arxiv.org/abs/1302.1204}
  {arXiv:1302.1204 [astro-ph.HE]} \BibitemShut {NoStop}%
\bibitem [{\citenamefont {Mahmoodifar}\ and\ \citenamefont
  {Strohmayer}(2017)}]{Mahmoodifar:2017ccc}%
  \BibitemOpen
  \bibfield  {author} {\bibinfo {author} {\bibfnamefont {S.}~\bibnamefont
  {Mahmoodifar}}\ and\ \bibinfo {author} {\bibfnamefont {T.}~\bibnamefont
  {Strohmayer}},\ }\href@noop {} {\bibfield  {journal} {\bibinfo  {journal}
  {Astrophys. J.}\ }\textbf {\bibinfo {volume} {840}},\ p.~\bibinfo {pages}
  {94} (\bibinfo {year} {2017})},\ \Eprint {http://arxiv.org/abs/1705.06780}
  {arXiv:1705.06780 [astro-ph.HE]} \BibitemShut {NoStop}%
\bibitem [{\citenamefont {Haskell}, \citenamefont {Glampedakis},\ and\
  \citenamefont {Andersson}(2014)}]{Haskell:2013hja}%
  \BibitemOpen
  \bibfield  {author} {\bibinfo {author} {\bibfnamefont {B.}~\bibnamefont
  {Haskell}}, \bibinfo {author} {\bibfnamefont {K.}~\bibnamefont
  {Glampedakis}}, \ and\ \bibinfo {author} {\bibfnamefont {N.}~\bibnamefont
  {Andersson}},\ }\href {\doibase 10.1093/mnras/stu535} {\bibfield  {journal}
  {\bibinfo  {journal} {Mon. Not. Roy. Astron. Soc.}\ }\textbf {\bibinfo
  {volume} {441}},\ \unskip\ \bibinfo {pages} {1662--1668} (\bibinfo {year}
  {2014})},\ \Eprint {http://arxiv.org/abs/1307.0985} {arXiv:1307.0985
  [astro-ph.SR]} \BibitemShut {NoStop}%
\bibitem [{\citenamefont {Alford}, \citenamefont {Han},\ and\ \citenamefont
  {Schwenzer}(2015)}]{Alford:2014jha}%
  \BibitemOpen
  \bibfield  {author} {\bibinfo {author} {\bibfnamefont {M.~G.}\ \bibnamefont
  {Alford}}, \bibinfo {author} {\bibfnamefont {S.}~\bibnamefont {Han}}, \ and\
  \bibinfo {author} {\bibfnamefont {K.}~\bibnamefont {Schwenzer}},\ }\href
  {\doibase 10.1103/PhysRevC.91.055804} {\bibfield  {journal} {\bibinfo
  {journal} {Phys. Rev. C}\ }\textbf {\bibinfo {volume} {91}},\ p.\ \bibinfo
  {pages} {055804} (\bibinfo {year} {2015})},\ \Eprint
  {http://arxiv.org/abs/1404.5279} {arXiv:1404.5279 [astro-ph.SR]} \BibitemShut
  {NoStop}%
\bibitem [{\citenamefont {Steiner}, \citenamefont {Hempel},\ and\ \citenamefont
  {Fischer}(2013)}]{Steiner:2012rk}%
  \BibitemOpen
  \bibfield  {author} {\bibinfo {author} {\bibfnamefont {A.~W.}\ \bibnamefont
  {Steiner}}, \bibinfo {author} {\bibfnamefont {M.}~\bibnamefont {Hempel}}, \
  and\ \bibinfo {author} {\bibfnamefont {T.}~\bibnamefont {Fischer}},\ }\href
  {\doibase 10.1088/0004-637X/774/1/17} {\bibfield  {journal} {\bibinfo
  {journal} {Astrophys. J.}\ }\textbf {\bibinfo {volume} {774}},\ p.~\bibinfo
  {pages} {17} (\bibinfo {year} {2013})},\ \Eprint
  {http://arxiv.org/abs/1207.2184} {arXiv:1207.2184 [astro-ph.SR]} \BibitemShut
  {NoStop}%
\bibitem [{\citenamefont {Lattimer}\ and\ \citenamefont
  {Schutz}(2005)}]{Lattimer:2004nj}%
  \BibitemOpen
  \bibfield  {author} {\bibinfo {author} {\bibfnamefont {J.~M.}\ \bibnamefont
  {Lattimer}}\ and\ \bibinfo {author} {\bibfnamefont {B.~F.}\ \bibnamefont
  {Schutz}},\ }\href {\doibase 10.1086/431543} {\bibfield  {journal} {\bibinfo
  {journal} {Astrophys. J.}\ }\textbf {\bibinfo {volume} {629}},\ \unskip\
  \bibinfo {pages} {979--984} (\bibinfo {year} {2005})},\ \Eprint
  {http://arxiv.org/abs/astro-ph/0411470} {arXiv:astro-ph/0411470 [astro-ph]}
  \BibitemShut {NoStop}%
\bibitem [{\citenamefont {Watts}\ \emph {et~al.}(2015)\citenamefont {Watts}
  \emph {et~al.}}]{Watts:2014tja}%
  \BibitemOpen
  \bibfield  {author} {\bibinfo {author} {\bibfnamefont {A.}~\bibnamefont
  {Watts}} \emph {et~al.},\ }\href@noop {} {\bibfield  {journal} {\bibinfo
  {journal} {PoS}\ }\textbf {\bibinfo {volume} {AASKA14}},\ p.\ \bibinfo
  {pages} {043} (\bibinfo {year} {2015})},\ \Eprint
  {http://arxiv.org/abs/1501.00042} {arXiv:1501.00042 [astro-ph.SR]}
  \BibitemShut {NoStop}%
\bibitem [{\citenamefont {Danielewicz}, \citenamefont {Lacey},\ and\
  \citenamefont {Lynch}(2002)}]{Danielewicz:2002pu}%
  \BibitemOpen
  \bibfield  {author} {\bibinfo {author} {\bibfnamefont {P.}~\bibnamefont
  {Danielewicz}}, \bibinfo {author} {\bibfnamefont {R.}~\bibnamefont {Lacey}},
  \ and\ \bibinfo {author} {\bibfnamefont {W.~G.}\ \bibnamefont {Lynch}},\
  }\href {\doibase 10.1126/science.1078070} {\bibfield  {journal} {\bibinfo
  {journal} {Science}\ }\textbf {\bibinfo {volume} {298}},\ \unskip\ \bibinfo
  {pages} {1592--1596} (\bibinfo {year} {2002})},\ \Eprint
  {http://arxiv.org/abs/nucl-th/0208016} {arXiv:nucl-th/0208016 [nucl-th]}
  \BibitemShut {NoStop}%
\end{thebibliography}%

\end{document}